\documentclass[conference,10pt]{IEEEtran}
\usepackage{epsfig,epsf,rotating,setspace,latexsym,amsmath,amssymb,amsfonts,bm,theorem,subfigure,epstopdf,cite,authblk,bbm,color,xcolor,mathtools,soul}
\usepackage{algorithm}
\usepackage[noend]{algpseudocode}

\algrenewcommand\algorithmicforall{\textbf{foreach}}
\algrenewcommand\algorithmicindent{.8em}

\newtheorem{theorem}{Theorem}
\newtheorem{lemma}{Lemma}

\newenvironment{Proof}[1]{\medskip\par\noindent{\bf Proof:\,}\,#1}{{\mbox{\,$\blacksquare$}\par}}

\IEEEoverridecommandlockouts
\allowdisplaybreaks

\begin{document}
 
\title{Age of Gossip in Ring Networks in the \\ Presence of Jamming Attacks }
 
\author{Priyanka Kaswan \qquad Sennur Ulukus\\
        \normalsize Department of Electrical and Computer Engineering\\
        \normalsize University of Maryland, College Park, MD 20742\\
        \normalsize  \emph{pkaswan@umd.edu} \qquad \emph{ulukus@umd.edu}}
 
\maketitle

\begin{abstract}
We consider a system with a source which maintains the most current version of a file, and a ring network of $n$ user nodes that wish to acquire the latest version of the file. The source gets updated with newer file versions as a point process, and forwards them to the user nodes, which further forward them to their neighbors using a memoryless gossip protocol. We study the average \emph{version age} of this network in the presence of $\tilde{n}$ jammers that disrupt inter-node communications. To this purpose, we construct an alternate system model of mini-rings and prove that the version age of the original model can be sandwiched between constant multiples of the version age of the alternate model. We show that when the number of jammers scales as a fractional power of the network size, i.e., $\tilde n= cn^\alpha$, the version age scales as $\sqrt{n}$ when $\alpha < \frac{1}{2}$, and as $n^{\alpha}$ when $\alpha \geq \frac{1}{2}$. As the version age of a ring network without any jammers scales as $\sqrt{n}$, our result implies that the version age with gossiping is robust against up to $\sqrt{n}$ jammers in a ring network.
\end{abstract}

\section{Introduction}

\begin{figure}[t]
\centerline{\includegraphics[width=0.5\linewidth]{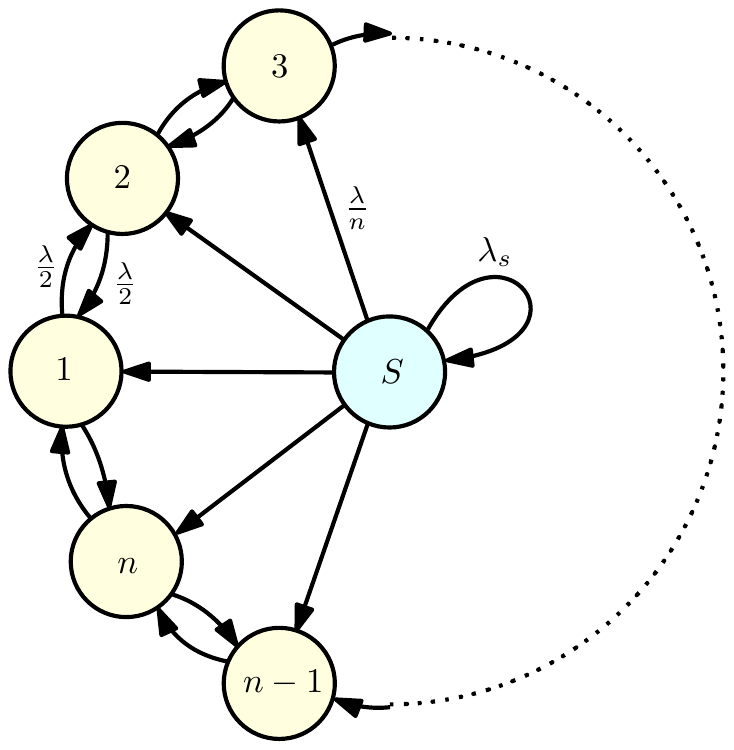}}
\caption{Ring network of $n$ nodes.}
\label{fig:ring_network}
\vspace*{-0.4cm}
\end{figure}

We investigate resilience of gossip-based information dissemination over networks used for timeliness purposes, against intentional jamming. We focus on a ring network, see Fig.~\ref{fig:ring_network}, where a source node which keeps the latest version of a file updates $n$ nodes placed on a ring with equal update rates $\frac{\lambda}{n}$. The source itself is updated with a rate $\lambda_s$. The nodes on the ring update their neighbors with equal update rates of $\frac{\lambda}{2}$. Without any jammers, the version age of a node in this network scales as $\sqrt{n}$ with the network size $n$ \cite{Yates21gossip, baturalp21comm_struc}. We address the following questions: If there are $\tilde{n}$ jammers in this network, how high can they drive the version age? What are the worst and least-harm jammer configurations? How many jammers does it take to drive the version age to order $n$, the version age with no gossiping among the nodes?

Gossip-based algorithms are used to disseminate information efficiently in large-scale networks with no central entity to coordinate exchange of information between users. Instead, users arbitrarily contact their neighbors and exchange information based on their local status, oblivious to the simultaneous dynamics of the network as a whole, thereby causing information to spread like a gossip/rumor. Gossip algorithms are simple and scalable, and have been applied in a wide range of contexts, such as, ad-hoc routing, distributed peer sampling, autonomic self management, data aggregation, and consensus. Gossiping is introduced in \cite{Demers1987EpidemicAF-short} to fix inconsistency in clearinghouse database servers, and since then has been studied extensively, e.g., \cite{Demers1987EpidemicAF-short, Minsky02cornellthesis, vocking2000, Pittel1987OnSA, deb2006AlgebraicGossip, devavrat2006, Sanghavi2007GossipFileSplit, amazondynamo-short, Cassandra, Yates21gossip, baturalp21comm_struc, Bastopcu21gossip, kaswan22slicingcoding}. For example, \cite{vocking2000} shows that a single rumor can be disseminated to $n$ nodes in $O(\log n)$ rounds, \cite{deb2006AlgebraicGossip} shows that using random linear coding (RLC) $n$ messages can be disseminated to $n$ nodes in $O(n)$ time in fully connected networks, \cite{devavrat2006} further extends this result to arbitrarily connected graphs, and  \cite{Sanghavi2007GossipFileSplit} presents an improved dissemination time by dividing files into $k$ pieces.

However, data in realistic systems is not static; it keeps changing asynchronously over time as new information becomes available. For example, distributed databases like Amazon DynamoDB \cite{amazondynamo-short} and Apache Cassandra \cite{Cassandra} use gossiping for real-time peer discovery and metadata propagation. In Cassandra, cluster metadata at each node is stored in endpoint state which tracks the version number or timestamp of the data. During a single gossip exchange between two nodes, the version number of the data at the two nodes is compared, and the node with older version number discards its data in favor of the more up-to-date data present at the other node. Hence, a specific information may get discarded or lost in the network before it can reach all nodes of the network. This renders the choice of total dissemination time as a performance metric inadequate, and in such networks, the \emph{age of information} at the nodes may prove to be a more suitable performance metric. 

Age of information has been studied in a range of contexts \cite{Kosta17agesurvey, Sun19agesurvey, yates21agesurvey}. In this paper, we use \emph{version} age of information metric \cite{Yates21gossip, Eryilmaz21, bastopcu20_google} together with exponential inter-update times as used previously in \cite{bastopcu20_google, Yates17sqrt, bastopcu2020LineNetwork, kaswan_isit2021}. Version age tracks the difference between the version number of the latest file at a node and the current version prevailing at the source. Gossip networks have been studied from an age of information point of view in \cite{Yates21gossip, baturalp21comm_struc, Bastopcu21gossip, kaswan22slicingcoding}, where \cite{Yates21gossip} derives a recursion to find the version age in arbitrary networks and characterizes the version age scaling in fully connected graphs, \cite{baturalp21comm_struc} proves the version age for ring networks and improves the version age scaling by introducing clustering, \cite{Bastopcu21gossip} derives analogous results for the binary freshness metric, and \cite{kaswan22slicingcoding} improves version age scaling using file slicing and network coding.

In this work, we focus on the version age in the presence of jammers  \cite{Nguyen17interferencegame, Garnaev19jamming,Xiao18jamming, Banerjee22adversary, Banerjee22game} for gossip networks. A jammer is a malicious entity that disrupts communication between two nodes, say by jamming the channel with noise. The jammer can also be a proxy for communication link failure, network partitioning, network congestion or information corruption during transfer, prevalent in distributed networks. Several works have characterized the effect of adversarial interference on gossip networks for the total dissemination time \cite{Augustine16_adversaries, Georgiou08_complexitygossip}. 

In this paper, we initiate a study of adversarial robustness of gossip networks from an age of information perspective. As an initial work in this direction, we focus on characterizing the impact of number of jammers on the version age scaling of gossip based \emph{ring networks} shown in Fig.~\ref{fig:ring_network}. We show that when the number of jammers $\tilde{n}$ scales as a fractional power of network size $n$, i.e., $\tilde n= cn^\alpha$, the average version age scales with a lower bound $\Omega(\sqrt{n})$ and an upper bound $O(\sqrt{n})$ when $\alpha \in \left[0,\frac{1}{2}) \right.$, and with a lower bound $\Omega(n^{\alpha})$ and an upper bound $O(n^{\alpha})$ when $\alpha \in \left[\frac{1}{2},1\right]$. To this purpose, we construct an alternate system model of mini-rings (see Fig.~\ref{fig:best_worst_jammer_positions}) and prove that the version age of the original model can be sandwiched between constant multiples of the version age of the alternate model. Along the way, we consider average version age in line networks (see Fig.~\ref{fig:line_network_model}) and prove structural results. 

\section{System Model and Preliminaries} \label{sec:sys_model}

The system model consists of a source, which maintains the most up-to-date version of a file and a large network of $n$ user nodes arranged on a ring (Fig.~\ref{fig:ring_network}), that wish to acquire the latest version of the file. The source is updated with newer file versions with exponential inter-update times with rate $\lambda_s$. The source forwards the current file to each user node with exponential inter-update times with rate $\frac{\lambda}{n}$, and further, each user node sends updates to each of its two  nearest neighbors with exponential inter-update times with rate $\frac{\lambda}{2}$. In addition, the system is faced with the presence of $\tilde{n}$ jammers which jam, i.e., cut, inter-node links, thereby disrupting any communication between the nodes connected by these links. 

When two jammers try to cut the same link, they will together be considered as a single jammer. That is, each jammer in our model is assumed to cut a distinct inter-node link. At time $t$, if $N_i(t)$ is the latest version of a file available at user node $i$ and $N(t)$ is the current version prevailing at the source, then the instantaneous version age at node $i$ at time $t$ is $\Delta_i(t)=N(t)-N_i(t)$. The long-term expected version age of node $i$ is given by $\Delta_i=\lim_{t \to \infty} \mathbb{E}[\Delta_i(t)]$. For a set $S$ of nodes, $\Delta_S(t)=\min_{i\in S}\Delta_i(t)$, and $\Delta_S=\lim_{t \to \infty} \mathbb{E}[\Delta_S(t)]$.

\begin{figure}[t]
\centerline{\includegraphics[width=0.73\linewidth]{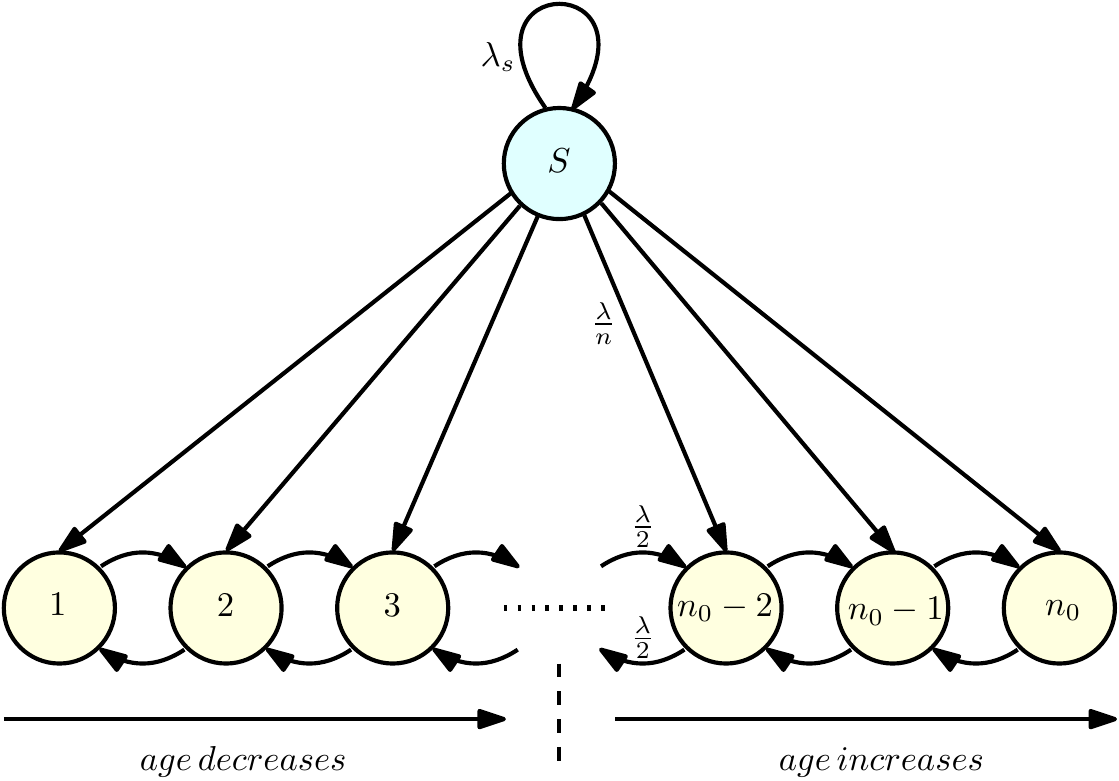}}
\caption{Line network model with $n_0$ nodes.}
\label{fig:line_network_model}
\vspace*{-0.4cm}
\end{figure}

When multiple adversaries cut inter-node communication links in this symmetric ring, the ring network is dismembered into a collection of isolated groups of nodes, where each group has the structure of a \emph{line network} shown in Fig.~\ref{fig:line_network_model}. The age of nodes in each such group are no longer statistically identical, owing to disappearance of circular symmetry. In this respect, we begin by examining the spatial variation of version age over a line network of $n_0$ nodes in the next section. 

\section{Version Age in a Line Network} \label{sec:age_progress_line}

Consider the line network of $n_0$ nodes as shown in Fig.~\ref{fig:line_network_model}. In Theorem~\ref{thm:age_variation_line} below, we show that the expected version age in this network is highest at the corner nodes and decreases towards the center. Intuitively, this is due to the fact that corner nodes are updated less frequently as they are connected with only one inter-node link, and thus, the exchanges towards the network corners are based on relatively staler file versions. Here, superscript $\ell(n_0)$ denotes a line network of size $n_0$.

\begin{theorem}\label{thm:age_variation_line}
    In a line network, $\Delta^{\ell(n_0)}_{i+1} \leq \Delta^{\ell(n_0)}_i$,  $i\leq\frac{n_0}{2}$. 
\end{theorem}

\begin{figure}[t]
\centerline{\includegraphics[width=1\linewidth]{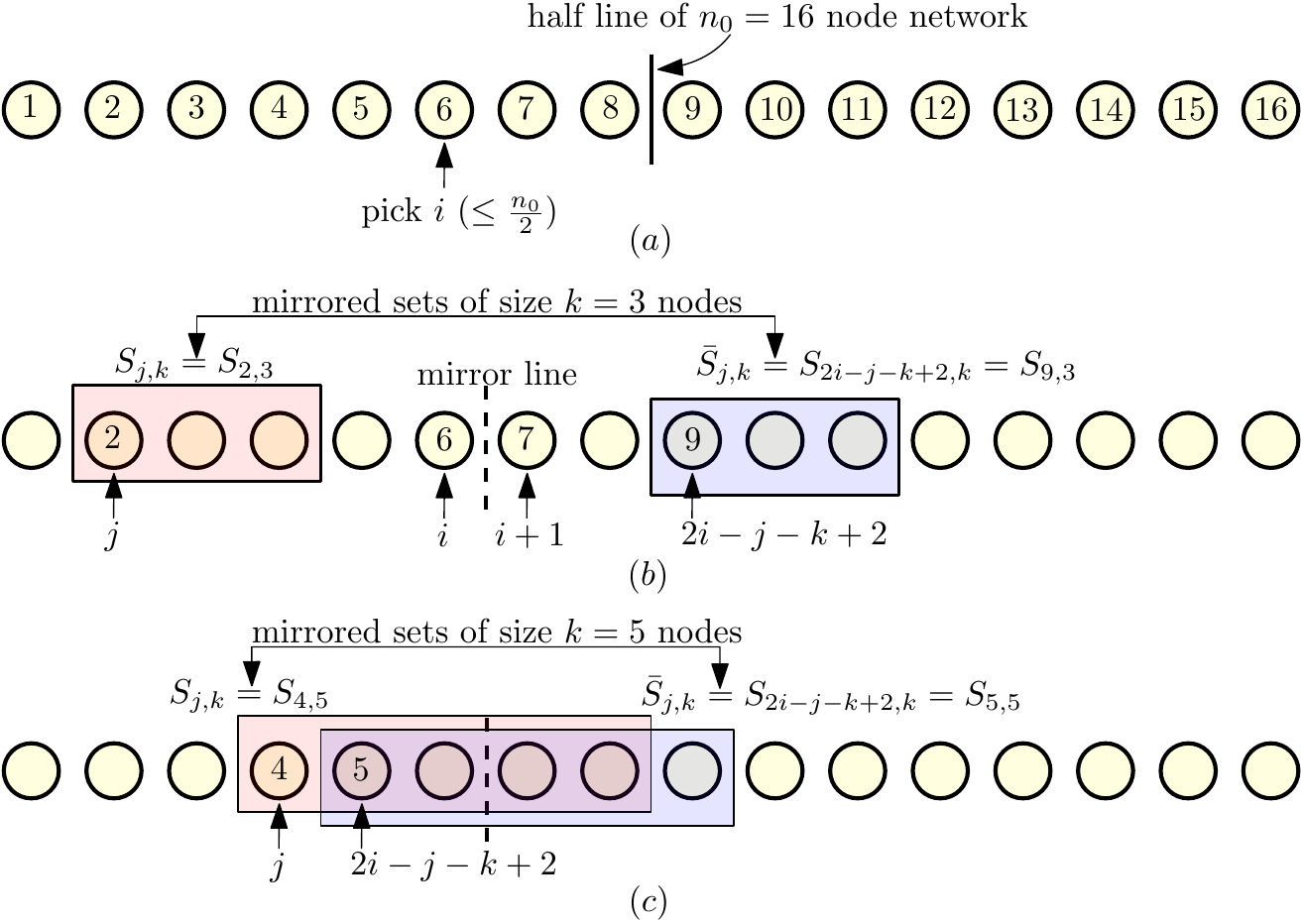}}
\caption{$S_{j,k}$ (pink blocks) and $\bar{S}_{j,k}=S_{2i-j-k+2,k}$ (blue blocks) positioned symmetrically about the dotted line between nodes $i$ and $i+1$.}
\label{fig:line_network_blocks}
\vspace*{-0.4cm}
\end{figure}

\begin{Proof}
Fix the $i$ in the statement of the theorem. Let $S_{j,k}=\{j,\ldots,j+k-1\}$ denote a set of $k$ contiguous nodes, beginning with node $j$, in a size $n_0$ line network, where $j+k-1\leq n_0$, see Fig.~\ref{fig:line_network_blocks}. We use $\Delta^{\ell(n_0)}_{j,k}$ to denote $\Delta^{\ell(n_0)}_{S_{j,k}}$, i.e., replace the set with its indices. Define $\bar{S}_{j,k}$ as the mirror image of set $S_{j,k}$ about the dotted line between nodes $i$ and $i+1$. Note that $\bar{S}_{j,k}=S_{2i-j-k+2,k}$; see Fig.~\ref{fig:line_network_blocks} for examples. Similarly, we use $\bar{\Delta}^{\ell(n_0)}_{j,k}$ to denote $\Delta^{\ell(n_0)}_{\bar{S}_{j,k}}$. We will consider sets $S_{j,k}$ where majority of the elements of $S_{j,k}$ lie to the left of node $i$, i.e., $j\leq i+1-\frac{k}{2}$, and prove that $\bar{\Delta}^{\ell(n_0)}_{j,k}\leq \Delta^{\ell(n_0)}_{j,k}$. Then, taking $k=1$ (size one set) with $j=i$ gives the desired result.  

We provide a proof by induction, beginning with the case $k=2i$, where $j=1$ is the only value that meets the condition $j\leq i+1-\frac{k}{2}$. This case gives $\bar{S}_{j,k} = S_{j,k}$ and $\bar{\Delta}^{\ell(n_0)}_{j,k}=\Delta^{\ell(n_0)}_{j,k}$ which satisfies the claim. Next, we assume that the claim holds for some $k$, i.e., $\bar{\Delta}^{\ell(n_0)}_{j,k}\leq \Delta^{\ell(n_0)}_{j,k}$ for all $j\leq i+1-\frac{k}{2}$, and prove it for $k-1$. To relate version ages of size $k$ and $k-1$ sets, we apply \cite[Thm.~1]{Yates21gossip} to obtain the version age of $S_{j,k-1}$
\begin{align}
\Delta^{\ell(n_0)}_{j,k-1}= \begin{cases} \label{eqn:k-1_block_age_trans_cases}
\frac{\lambda_s+\frac{\lambda}{2}\Delta^{\ell(n_0)}_{j,k}+\frac{\lambda}{2}\Delta^{\ell(n_0)}_{j-1,k}}{\frac{(k-1)\lambda}{n}+\lambda}, & j> 1\\
\frac{\lambda_s+\frac{\lambda}{2}\Delta^{\ell(n_0)}_{j,k}}{\frac{(k-1)\lambda}{n}+\frac{\lambda}{2}},& j=1
\end{cases}
\end{align}
and for its mirrored set $\bar{S}_{j,k-1}$ as
\begin{align} \label{eqn:k-1_mirrorblock_age_trans}
    \bar{\Delta}^{\ell(n_0)}_{j,k-1}=\frac{\lambda_s+\frac{\lambda}{2} \bar{\Delta}^{\ell(n_0)}_{j,k} +\frac{\lambda}{2} \bar{\Delta}^{\ell(n_0)}_{j-1,k}}{\frac{(k-1)\lambda}{n}+\lambda}
\end{align}
Since $\bar{S}_{j,k-1} \subset \bar{S}_{j-1,k}$, we have $\bar{\Delta}^{\ell(n_0)}_{j,k-1} \geq  \bar{\Delta}^{\ell(n_0)}_{j-1,k}$ as an extended feasible region provides a lower minimum. Using this inequality to eliminate the last term in (\ref{eqn:k-1_mirrorblock_age_trans}) gives
\begin{align} \label{eqn:k-1_mirrorblock_age_trans_inequality}
    \bar{\Delta}^{\ell(n_0)}_{j,k-1} \leq \frac{\lambda_s+\frac{\lambda}{2}\bar{\Delta}^{\ell(n_0)}_{j,k}}{\frac{(k-1)\lambda}{n}+\frac{\lambda}{2}}
\end{align}

We wish to prove the claim for $k-1$, i.e., $\bar{\Delta}^{\ell(n_0)}_{j,k-1} \leq \Delta^{\ell(n_0)}_{j,k-1}$ under the condition $j\leq i+1-\frac{(k-1)}{2}$. If $j=i+1-\frac{(k-1)}{2}$, then we get $j=2i-j-k+3$, giving $\bar{S}_{j,k-1} = S_{j,k-1}$ and $\bar{\Delta}^{\ell(n_0)}_{j,k-1} = \Delta^{\ell(n_0)}_{j,k-1}$ which satisfies the claim. For other $j$, as $j$ must be integer valued, $j\leq i+1-\frac{(k-1)}{2}$ is automatically implied from  $j\leq i+1-\frac{k}{2}$. Consequently, since the claim is assumed to hold for $k$, we have $\bar{\Delta}^{\ell(n_0)}_{j,k}\leq \Delta^{\ell(n_0)}_{j,k}$ and $\bar{\Delta}^{\ell(n_0)}_{j-1,k}\leq \Delta^{\ell(n_0)}_{j-1,k}$. Now, comparing the terms in (\ref{eqn:k-1_block_age_trans_cases}) and (\ref{eqn:k-1_mirrorblock_age_trans}) for case $j>1$, and (\ref{eqn:k-1_block_age_trans_cases}) and (\ref{eqn:k-1_mirrorblock_age_trans_inequality}) for case $j=1$ gives $\bar{\Delta}^{\ell(n_0)}_{j,k-1} \leq \Delta^{\ell(n_0)}_{j,k-1}$, and thus, the claim holds for $k-1$. Finally, setting $k=1$ and $j=i$, we have $\Delta^{\ell(n_0)}_{i+1}=\Delta^{\ell(n_0)}_{i+1,1}=\bar{\Delta}^{\ell(n_0)}_{i,1}\leq \Delta^{\ell(n_0)}_{i,1}=\Delta^{\ell(n_0)}_i$, which completes the proof.
\end{Proof}

\section{Bounds on Age of Line Network} \label{sec:bounds_line}

\begin{figure}[t]
\centerline{\includegraphics[width=0.8\linewidth]{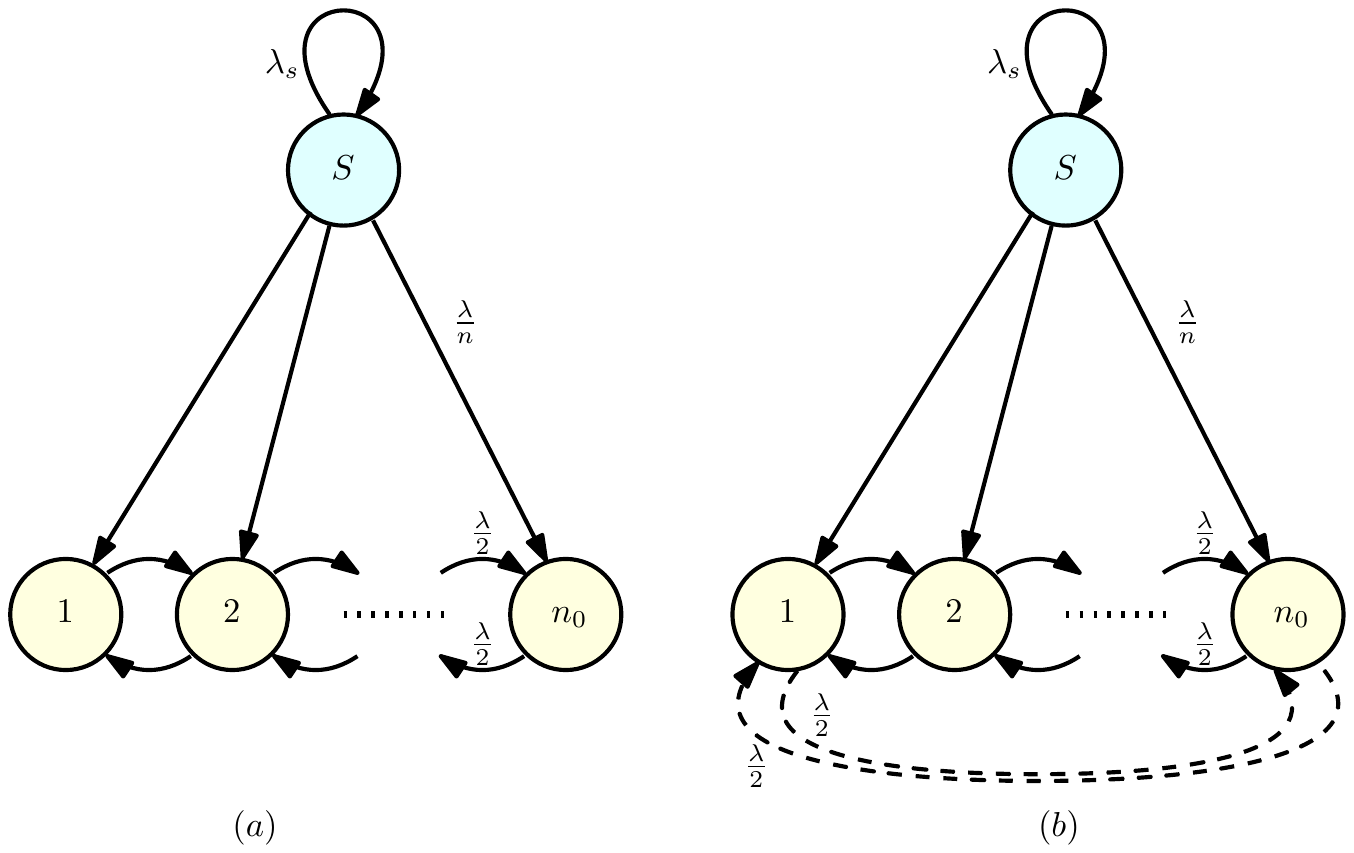}}
\caption{(a) A line network of size $n_0$. (b) A ring network of size $n_0$. Compared to (a), an extra link is introduced between end nodes bringing radial symmetry.}
\label{fig:line_to_miniring}
\vspace*{-0.4cm}
\end{figure}

Here, we bound $\Delta^{\ell(n_0)}_{i}$, age in a line network of size $n_0$ with $\Delta^{r(n_0)}_{i}$, age in a ring network of size $n_0$, see Fig.~\ref{fig:line_to_miniring}. 

\subsection{Lower Bound} \label{subsec:lowerbound_line}

Heuristically, due to the presence of an additional link compared to the line network, the ring network will have more age-conformed transitions, and therefore, improved age at the nodes. Mathematically, the recursive equation in \cite[Thm.~1]{Yates21gossip} is identical in the two networks for every subset of nodes $S_{j,k} \in \{2,\ldots,n_0-1\}$ that excludes corner nodes $1$ and $n_0$. 

Further, for subsets $S_{1,k}$ which include corner node $1$, similar to  (\ref{eqn:k-1_block_age_trans_cases}) for case $j=1$, we have
\begin{align} \label{eqn:line_end_node}
    \Delta^{\ell(n_0)}_{1,k}=\frac{\lambda_s+\frac{\lambda}{2}\Delta^{\ell(n_0)}_{1,k+1}}{\frac{k\lambda}{n}+\frac{\lambda}{2}}
\end{align}
The corresponding equation for the ring, similar to (\ref{eqn:k-1_mirrorblock_age_trans})-(\ref{eqn:k-1_mirrorblock_age_trans_inequality}), is
\begin{align}
    \!\!\! \Delta^{r(n_0)}_{1,k}=\frac{\lambda_s+\frac{\lambda}{2}\Delta^{r(n_0)}_{1,k+1}+\frac{\lambda}{2}\Delta^{r(n_0)}_{n_0,k+1}}{\frac{k\lambda}{n}+\lambda}
    \leq \frac{\lambda_s+\frac{\lambda}{2}\Delta^{r(n_0)}_{1,k+1}}{\frac{k\lambda}{n}+\frac{\lambda}{2}} \label{eqn:ring_end_node}
\end{align}
where $S_{n_0,k+1}$ refers to the set $\{n_0,1,2,\ldots,k\}$, and hence, $S_{1,k}\subset S_{n_0,k+1}$ gives the inequality $\Delta^{r(n_0)}_{n_0,k+1} \leq \Delta^{r(n_0)}_{1,k}$.

Now, comparing (\ref{eqn:line_end_node}) and (\ref{eqn:ring_end_node}) gives $\Delta^{r(n_0)}_{1,k} \leq \Delta^{\ell(n_0)}_{1,k}$. By symmetry, $ \Delta^{r(n_0)}_{n_0+1-k,k} \leq \Delta^{\ell(n_0)}_{n_0+1-k,k}$ for subsets which include the other corner node $n_0$. Thus, $\Delta^{r(n_0)}_{j,k} \leq \Delta^{\ell(n_0)}_{j,k}$, for all $j,k$, and $\Delta^{r(n_0)}_{i} \leq \Delta^{\ell(n_0)}_{i}$, which gives us a lower bound on $\Delta^{\ell(n_0)}_{i}$.

\subsection{Upper Bound}

Note that due to the radial symmetry of the ring network, $\Delta^{r(n_0)}_i=\Delta^{r(n_0)}_1$, for all $i$. Hence, combining results of Sections~\ref{sec:age_progress_line}~and~\ref{subsec:lowerbound_line}, with bilateral symmetry of line network, 
\begin{align}
    \Delta^{r(n_0)}_{1} &\leq \Delta^{\ell(n_0)}_{\lceil\frac{n_0}{2}\rceil}=\Delta^{\ell(n_0)}_{\lfloor\frac{n_0}{2}\rfloor}\leq \ldots \nonumber \\
    &\ldots \leq\Delta^{\ell(n_0)}_{n_0-1}=\Delta^{\ell(n_0)}_{2}\leq \Delta^{\ell(n_0)}_{n_0}=\Delta^{\ell(n_0)}_{1} 
\end{align}
Hence, an upper bound on $\Delta^{\ell(n_0)}_{1}$ is an upper bound on $\Delta^{\ell(n_0)}_{i}$ for all $i$. Recursively applying \cite[Thm.~1]{Yates21gossip}, see also \cite[Lem.~2]{baturalp21comm_struc},
\begin{align}\label{eqn:ring_end_node_detailage}
    \Delta^{r(n_0)}_1=\frac{\lambda_s}{\lambda}\Bigg[\sum_{j=1}^{n_0-1}\prod_{k=1}^{j}\frac{1}{\frac{k}{n}+1}+\frac{1}{\frac{n_0}{n}}\prod_{k=1}^{n_0-1}\frac{1}{\frac{k}{n}+1}\Bigg]
\end{align}
and
\begin{align}
    \Delta^{\ell(n_0)}_1=&\frac{2\lambda_s}{\lambda} \Bigg[\sum_{j=1}^{n_0-1}\prod_{k=1}^{j}\frac{1}{\frac{k}{n/2}+1} +\frac{1}{\frac{n_0}{n/2}}\prod_{k=1}^{n_0-1}\frac{1}{\frac{k}{n/2}+1}\Bigg]\\
    \leq & \frac{\lambda_s}{\lambda}\Bigg[\sum_{j=1}^{n_0-1}\prod_{k=1}^{j}\frac{1}{\frac{k}{n}+1}+\frac{1}{\frac{n_0}{n}}\prod_{k=1}^{n_0-1}\frac{1}{\frac{k}{n}+1}\Bigg] \\
    =&2\Delta^{r(n_0)}_1 
\end{align}

Hence for each node $i$, 
\begin{align}
\Delta^{r(n_0)}_1\leq \Delta^{\ell(n_0)}_i \leq 2\Delta^{r(n_0)}_1
\end{align}
Since $\Delta^{\ell(n_0)}_i$ is sandwiched between constant multiples of $\Delta^{r(n_0)}_1$, they both scale similarly when $n$ is large. Intuitively, when the number of nodes is large, the effect of one additional link in the \emph{line} versus \emph{ring} models becomes insignificant.

\begin{figure}[t]
\centerline{\includegraphics[width=1\linewidth]{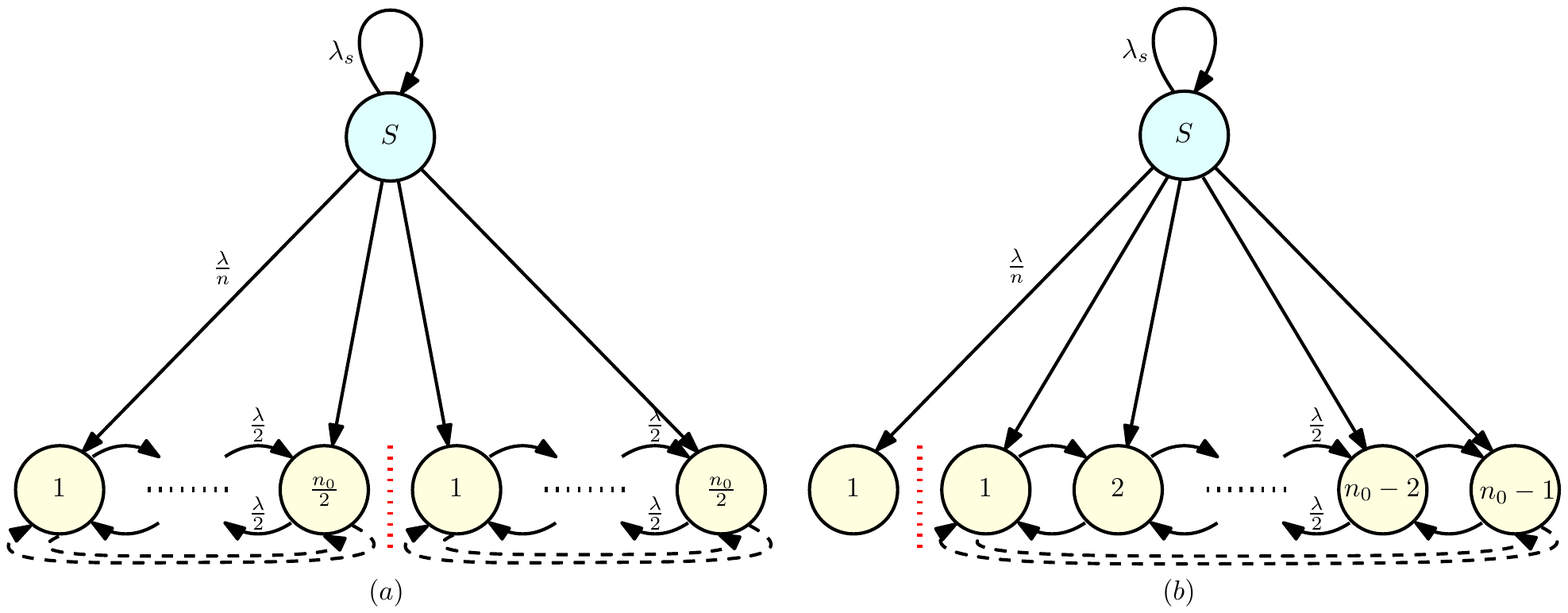}}
\vspace*{-0.2cm}
\caption{Jammer position on line network (a) most favorable, (b) most harmful.}
\label{fig:best_worst_rings}
\vspace*{-0.4cm}
\end{figure}

\section{Jammer Positioning for Age Deterioration} \label{subsec:position_jammers}

To characterize the total age of the system, we define $\Delta^{\ell(n_0)}=\sum_{i=1}^{n_0}\Delta^{\ell(n_0)}_i$ and $\Delta^{r(n_0)}=\sum_{i=1}^{n_0}\Delta^{r(n_0)}_i$, the sum of version ages in size $n_0$ line and ring networks, respectively. 

\begin{lemma}\label{lemma:ring_age_difference}
$\Delta^{r(n_0)} -\Delta^{r(n_0+1)}$ decreases with increase in $n_0$.
\end{lemma}

\begin{Proof}
We have $\Delta^{r(n_0)}=n_0 \Delta^{r(n_0)}_1$ due to the radial symmetry of a ring. Hence, by obtaining expressions for $\Delta^{r(n_0)}_1$ and $\Delta^{r(n_0+1)}_1$ from (\ref{eqn:ring_end_node_detailage}), we get
\begin{align}
    \Delta^{r(n_0)}-\Delta^{r(n_0+1)}=& n_0\Delta^{r(n_0)}_1-(n_0+1)\Delta^{r(n_0+1)}_1 \\
    =&\frac{\lambda_s}{\lambda}\Bigg[-\sum_{j=1}^{n_0}\prod_{k=1}^{j}\frac{1}{\frac{k}{n}+1}\Bigg] 
\end{align}
which decreases with increasing $n_0$.
\end{Proof}

Next, we consider the problem where a single jammer cuts a single link in the line network model of Fig.~\ref{fig:line_network_model}. Let the jammer cut the link between nodes $m$ and $m+1$, effectively converting the line network of size $n_0$ into two smaller line networks of sizes $m$ and $n_0-m$. The age of the resulting system is $\Delta^{\ell(n_0)(m)}=\Delta^{\ell(m)} +\Delta^{\ell(n_0-m)}$. We will consider the mini-ring approximation to the resulting two line networks, and assume that the end points of the individual line networks are connected to form mini-rings as shown in Fig.~\ref{fig:best_worst_rings}. The age of this network with mini-rings is $\Delta^{r(n_0)(m)}=\Delta^{r(m)} +\Delta^{r(n_0-m)}$. From Section~\ref{sec:bounds_line}, the total ages of the actual dismembered line network and the mini-ring approximation are related as $\Delta^{r(n_0)(m)} \leq \Delta^{\ell(n_0)(m)}\leq 2\Delta^{r(n_0)(m)}$. Next, as an approximation to finding $m$ that maximizes (worst case jammer) and minimizes (most favorable jammer) $\Delta^{\ell(n_0)(m)}$, we will find $m$ that maximizes/minimizes $\Delta^{r(n_0)(m)}$ instead. We show in Theorem~\ref{thm:best_position_for_jammer} below that most harmful jammer cuts the link that separates the node at the corner, and the most favorable jammer cuts the center link, as shown in Fig.~\ref{fig:best_worst_rings}.

\begin{theorem} \label{thm:best_position_for_jammer}
In a ring, $\Delta^{r(n_0)(m+1)}\leq \Delta^{r(n_0)(m)}$, $m \leq \frac{n_0}{2}$.
\end{theorem}

\begin{Proof}
We have 
\begin{align}
    &\Delta^{r(n_0)(m)}-\Delta^{r(n_0)(m+1)} \nonumber\\
    &= \left[\Delta^{r(m)} +\Delta^{r(n_0-m)}\right]- \left[\Delta^{r(m+1)} +\Delta^{r(n_0-m-1)}\right] \\
    &= \left[\Delta^{r(m)} -\Delta^{r(m+1)}\right] - \left[\Delta^{r(n_0-m-1)} -\Delta^{r(n_0-m)}\right]\\
    &\geq 0 
\end{align}
where the last inequality follows from Lemma~\ref{lemma:ring_age_difference}, as $m<n_0-m-1$ since we are only considering $m\leq \frac{n_0}{2}$.
\end{Proof}

\begin{figure}[t]
\centerline{\includegraphics[width=0.8\linewidth]{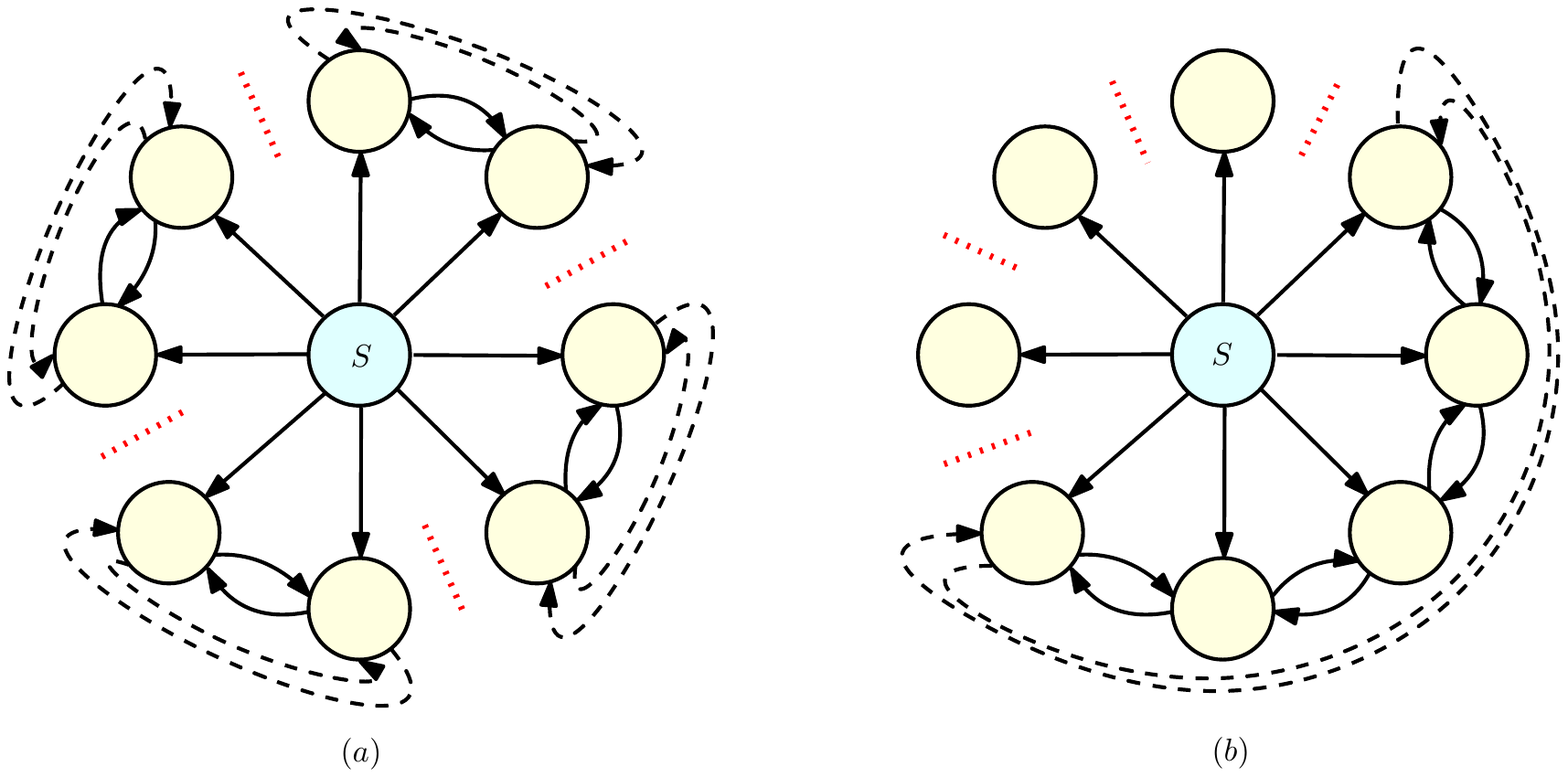}}
\vspace*{-0.2cm}
\caption{Jammer positions on a ring (a) most favorable, (b) most harmful.}
\label{fig:best_worst_jammer_positions}
\vspace*{-0.5cm}
\end{figure}

Consider the original ring network of $n$ nodes in Fig.~\ref{fig:ring_network} in the presence of $\tilde{n}$ jammers cutting $\tilde{n}$ communication links. This results in $\tilde{n}$ isolated line networks. Let $\Delta^\ell$ denote the average version age at in this dismembered ring, which is composed of $\tilde{n}$ line networks. Consider the alternate model, where all the $\tilde{n}$ line networks are replaced by their mini-ring versions, as shown in Fig.~\ref{fig:best_worst_jammer_positions}(a), and let $\Delta^r$ denote the average age of this alternate system. Then, from Section~\ref{sec:bounds_line}, we know that $\Delta^\ell$ is bounded by constant multiples of $\Delta^r$.

From Theorem~\ref{thm:best_position_for_jammer}, the least harmful positioning of jammers for $\Delta^r$ is the equidistant placement around the ring, as shown in Fig.~\ref{fig:best_worst_jammer_positions}(a), since we can keep switching a jammer's position untill it has equal size mini-rings on both sides. Likewise, the most detrimental positioning of jammers is if they cut adjacent links, as shown in Fig.~\ref{fig:best_worst_jammer_positions}(b), which results in $\tilde{n}-1$ isolated nodes and a single line network of $n-\tilde{n}+1$ nodes. This is because from Theorem~\ref{thm:best_position_for_jammer}, presence of two line networks of $i_1$ and $i_2$ nodes has lower age than presence of a single line network of $i_1+i_2-1$ nodes and an isolated node.

\section{Average Age of Ring Network with Jammers}

\begin{lemma}\label{lemma:gaussian_approx} (a)
$\sum_{j=1}^{n_0}\left[\prod_{k=1}^{j}\frac{1}{\frac{k}{n}+1}\right]=O(\sqrt{n})$. (b) If $n_0=\omega(\sqrt{n})$, then $\sum_{j=1}^{n_0}\left[\prod_{k=1}^{j}\frac{1}{\frac{k}{n}+1}\right]=\Omega(\sqrt{n})$.
\end{lemma}
\begin{Proof}
We use $\frac{x}{1+x} \leq \log(1+x) \leq x$ and $\sum_{k=1}^{j} \frac{k}{n} = \frac{j(j+1)}{2n}$ to bound $\sum_{k=1}^{j} \log \big( 1+\frac{k}{n} \big)$ and then exponentiate to obtain
\begin{align}\label{eqn:negative_exponent_bounds}
    e^{-\frac{j^2}{n}} \leq  \prod_{k=1}^{j} \frac{1}{1+\frac{k}{n}}  \leq e^{-\frac{j^2}{4n}}
\end{align}
Then, we sum over $j$ and use Riemann sums to obtain
\begin{align}
    \int_{\frac{1}{\sqrt{n}}}^{\frac{n_0}{\sqrt{n}}}e^{-\frac{t^2}{C}}dt \leq \frac{1}{\sqrt{n}}\sum_{j=1}^{n_0}  e^{-\frac{j^2}{Cn}} \leq \int_{0}^{\frac{n_0}{\sqrt{n}}}e^{-\frac{t^2}{C}}dt 
\end{align}
and note that $\int_{0}^{\infty}e^{-\frac{t^2}{C}}dt = \frac{\sqrt{C\pi}}{2}$ for a constant $C>0$.
\end{Proof}

\subsection{Lower Bound on System Age}

\begin{theorem} \label{thm:lowerbound}
For a ring network of $n$ nodes with $\tilde{n} =c n^{\alpha}$ jammers, where $\alpha \in [0,1]$ and $c$ is a scaling constant,
\begin{align}
\Delta^{\ell} = \begin{cases} 
\Omega(n^{\alpha}), & \alpha\geq \frac{1}{2}\\
\Omega(\sqrt{n}), & \alpha<\frac{1}{2}
\end{cases}
\end{align}
\end{theorem}

\begin{Proof}
Based on Section~\ref{subsec:position_jammers}, to lower bound $\Delta^r(\leq \Delta^{\ell})$, we consider the least detrimental model of Fig.~\ref{fig:best_worst_jammer_positions}(a), where each node is a part of $n_0=\frac{n}{\tilde n}=\frac{1}{c}n^{1-\alpha}$size mini-ring, and hence all nodes have identical average age giving $\Delta^r=\Delta^{r(n_0)}_1$.

For $\alpha<\frac{1}{2}$, using Lemma~\ref{lemma:gaussian_approx}(b) in (\ref{eqn:ring_end_node_detailage}) gives
\begin{align} \label{eqn:lowerbound_alpha3}
    \!\!\!\Delta^{r(n_0)}_1 \!\geq\! \frac{\lambda_s}{\lambda}\Bigg[\sum_{j=1}^{n_0-1}\!\prod_{k=1}^{j}\frac{1}{\frac{k}{n}+1}\!+\!\frac{n}{n_0}\!\prod_{k=1}^{n_0}\frac{1}{\frac{k}{n}+1}\Bigg] 
    \!\geq\!\Omega(\sqrt{n}) 
\end{align}
For $\alpha \geq \frac{1}{2}$, using (\ref{eqn:negative_exponent_bounds}) in (\ref{eqn:ring_end_node_detailage}) gives
\begin{align} \label{eqn:lowerbound_alpha8}
    \Delta^{r(n_0)}_1 \geq &\frac{\lambda_s}{\lambda}\left[0 +c n^{\alpha}e^{-\frac{n_0^2}{n}} \right] =\Omega(n^{\alpha})
\end{align}
where $e^{-\frac{n_0^2}{n}}= e^{-\frac{n^{-(2\alpha-1)}}{2c^2}}\to 1$ for large $n$.
\end{Proof}

\subsection{Upper Bound on System Age}
\begin{theorem} \label{thm:upperbound}
For a ring network of $n$ nodes with $\tilde{n} =c n^{\alpha}$ jammers, where $\alpha \in (0,1)$ and $c$ is a scaling constant, 
\begin{align}
\Delta^{\ell} = \begin{cases} 
O(n^{\alpha}), & \alpha\geq \frac{1}{2}\\
O(\sqrt{n}), & \alpha<\frac{1}{2}
\end{cases} 
\end{align}
\end{theorem}

\begin{Proof}
To upper bound $\Delta^r(\geq \frac{1}{2}\Delta^{\ell})$, we consider the most detrimental model of Fig.~\ref{fig:best_worst_jammer_positions}(b), which has $\tilde n -1$ isolated nodes and a line network of $n-\tilde{n} +1$ nodes. Then,
\begin{align} \label{eqn:average_worstcase_jammer}
    \Delta^r&= \frac{(\tilde n -1)\Delta^{r(1)}_1 + (n-\tilde n +1) \Delta^{r(n-\tilde{n} +1)}_1}{n} 
\end{align}
Since $\frac{\tilde n-1}{n} \leq \frac{\tilde n}{n}$ and $\frac{n-\tilde n+1}{n}=\frac{n(1-cn^{-(1-\alpha)}+\frac{1}{n})}{n} \leq 1$, using (\ref{eqn:ring_end_node_detailage}), (\ref{eqn:negative_exponent_bounds}) and Lemma~\ref{lemma:gaussian_approx}(a) in (\ref{eqn:average_worstcase_jammer}) gives
\begin{align} \label{eqn:upperbound}
    \Delta^r \leq& \frac{\tilde n}{n}\frac{\lambda_s n}{\lambda}+\frac{n-\tilde n+1}{n} \frac{\lambda_s}{\lambda} \left[ O(\sqrt{n}) + \frac{n c e^{-\frac{(n-\tilde n)^2}{4n}} }{n-\tilde n+1}\right] \\
    \leq&\frac{\lambda_s}{\lambda}\left[cn^{\alpha} + O(\sqrt{n}) + c \right] 
\end{align}
completing the proof.
\end{Proof}

\section{Numerical Results}

\begin{figure}[t]
 	\begin{center}
 	\subfigure[]{\includegraphics[width=0.49\linewidth]{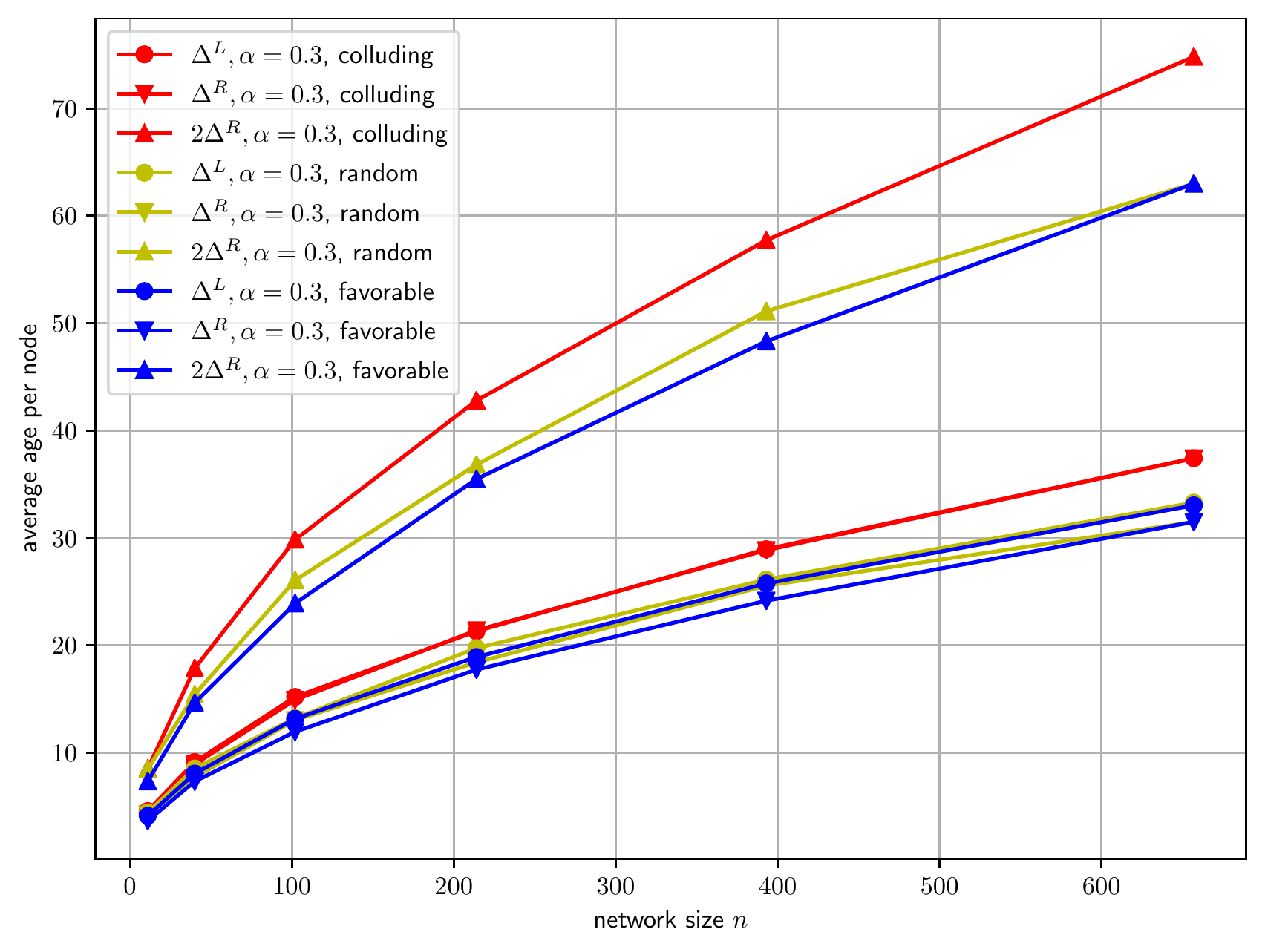}}
 	\subfigure[]{\includegraphics[width=0.49\linewidth]{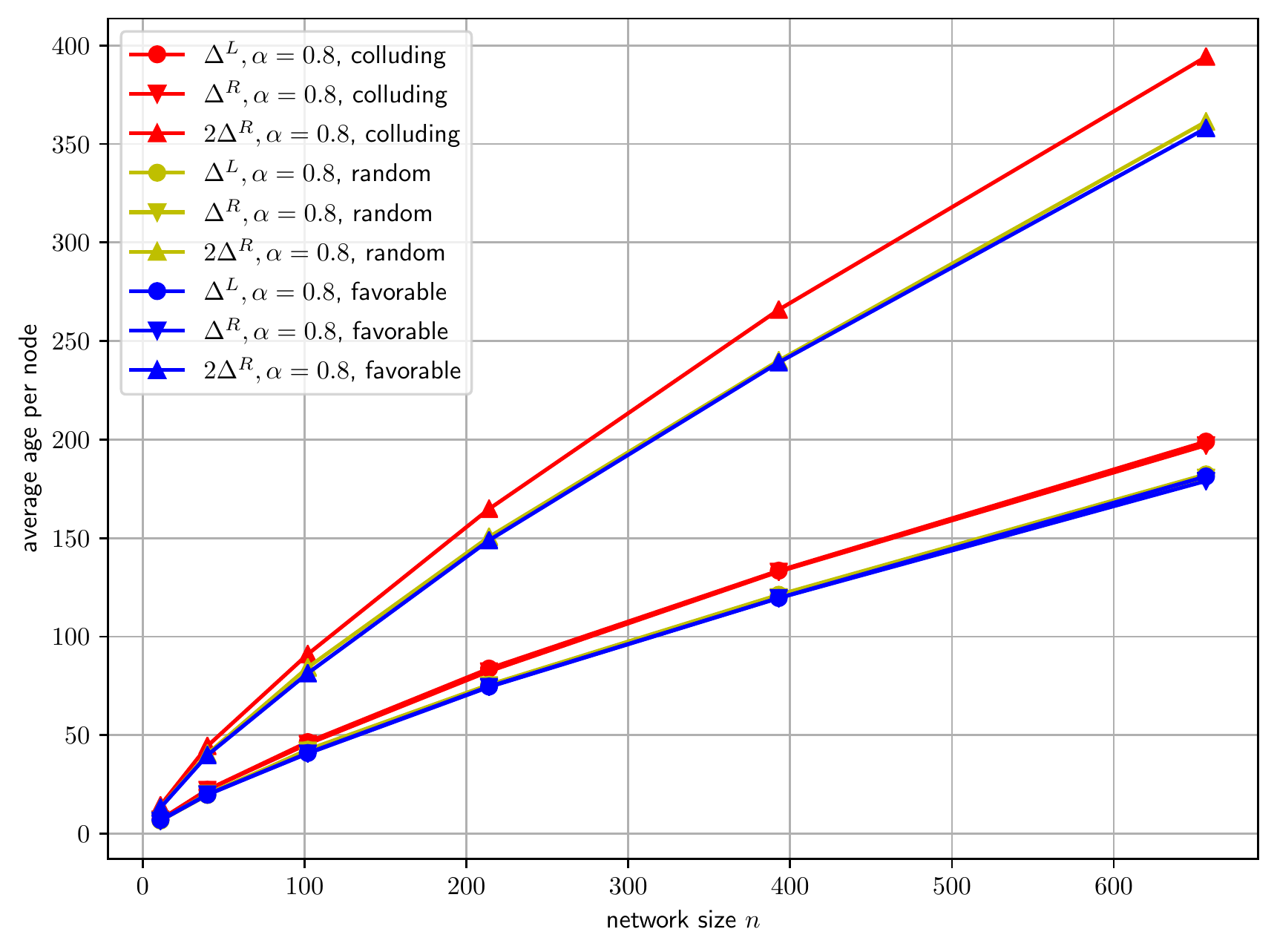}}
 	\end{center}
 	\vspace{-0.4cm}
 	\caption{(a) Average age with $n^{0.3}$. (b) Average age with $n^{0.8}$ jammers.}
 	\label{fig:average_age_with_alpha3}
 	\vspace{-0.65cm}
 \end{figure}

We now validate the aforementioned bounds by performing real-time gossip protocol simulations for both dismembered ring and its altered mini-ring model for three cases of jammer positions, a) most harmful (colluding), b) random, and c) least harmful (favorable) jammer positions. For $\tilde{n}= n^{\alpha}$ and $\frac{\lambda_s}{\lambda}=1$, Fig.~\ref{fig:average_age_with_alpha3}(a) and (b) show the average age plots as a function of the network size $n$ for $\alpha=0.3$ and $\alpha=0.8$, respectively. We observe that $\Delta^{\ell}$ is closely approximated by its lower bound $\Delta^r$ in all cases. Comparing $\alpha=0.3$ and $\alpha =0.8$ plots, the ages in the latter case increase steeply owing to the presence of larger number of jammers, and the graphs look almost linear because as $\alpha$ approaches $1$, $n^{\alpha}$ begins to appear linear. 

\bibliographystyle{unsrt}
\bibliography{ref_priyanka}

\end{document}